# The benefits and costs of explainable artificial intelligence in visual quality control: Evidence from fault detection performance and eye movements

Romy Müller[1], David F. Reindel[2], Yannick D. Stadtfeld[1]

[1] Faculty of Psychology, Chair of Engineering Psychology and Applied Cognitive Research, TUD Dresden University of Technology, Dresden, Germany

[2] Department of Child and Adolescent Psychiatry, Psychosomatics, and Psychotherapy, Uniklinikum Würzburg, Würzburg, Germany

Corresponding author:

Romy Müller

Chair of Engineering Psychology and Applied Cognitive Research

TUD Dresden University of Technology

Helmholtzstraße 10, 01069 Dresden, Germany

Email: romy.mueller@tu-dresden.de

Phone: +49 351 46335330

ORCID: 0000-0003-4750-7952

## Abstract

Visual inspection tasks often require humans to cooperate with AI-based image classifiers. To enhance this cooperation, explainable artificial intelligence (XAI) can highlight those image areas that have contributed to an AI decision. However, the literature on visual cueing suggests that such XAI support might come with costs of its own. To better understand how the benefits and cost of XAI depend on the accuracy of AI classifications and XAI highlights, we conducted two experiments that simulated visual quality control in a chocolate factory. Participants had to decide whether chocolate moulds contained faulty bars or not, and were always informed whether the AI had classified the mould as faulty or not. In half of the experiment, they saw additional XAI highlights that justified this classification. While XAI speeded up performance, its effects on error rates were highly dependent on (X)AI accuracy. XAI benefits were observed when the system correctly detected and highlighted the fault, but XAI costs were evident for misplaced highlights that marked an intact area while the actual fault was located elsewhere. Eye movement analyses indicated that participants spent less time searching the rest of the mould and thus looked at the fault less often. However, we also observed large interindividual differences. Taken together, the results suggest that despite its potentials, XAI can discourage people from investing effort into their own information analysis.

# 1 Introduction

Explainable artificial intelligence (XAI) can enhance human-technology cooperation by making the black box of artificial intelligence (AI) more transparent. For instance, decisions of image classifiers can be explained by highlighting those image areas that had an impact on the classification decision. In this way, humans should be able to judge whether the AI has made the right decision for the right reasons. This principle can be used to support visual inspection tasks, for instance in radiological image reading, luggage screening, industrial fault detection, or product quality control. However, the literature on visual cueing suggests that this type of support may come with its own risks: people might overrely on the XAI highlights instead of thoroughly analysing the image material, resulting in wrong decisions when the highlights are wrong. In the present study, we asked how the accuracy of AI decisions and XAI highlights moderates the influence of XAI on visual inspection processes and performance. Participants had to judge whether chocolate moulds contained faulty bars and were supported by a simulated AI system. This system either provided only indirect cues in the form of black box AI (i.e., indicating for each image whether the AI has detected a fault) or additional direct cues in the form of XAI (i.e., highlighting the presumed fault location). Before outlining our specific research questions, we will specify the role XAI can play during visual inspection, discuss what we can learn from the visual cueing literature, and point out what questions have remained unanswered so far.

## 1.1 Supporting Visual Inspection with XAI

Nowadays, powerful AI-based image classifiers can match or even surpass human abilities (e.g., Buetti-Dinh et al., 2019; Kshatri & Singh, 2023). At the same time, they make surprising errors in tasks that are trivial for humans (Geirhos et al., 2020). Accordingly, it seems promising to combine the capabilities of humans and AI during visual inspection to optimise overall system performance. For instance, an AI system could indicate whether it has classified an image as containing a target or not, while a human still is responsible for the final decision. With this type of support, the AI provides an indirect cue (Chavaillaz et al., 2018; Goh et al., 2005): it merely informs observers whether a target is present, but does not say where. Sometimes reliable indirect cues improve human detection performance (e.g., Boskemper et al., 2022), but at other times they do not, because people keep looking for targets by themselves (e.g., Goh et al., 2005). In principle, it can be considered a good thing when people do not blindly rely on AI without cross-checking its decisions. However, this offsets the potential benefits of AI, because people still perform the entire task. In doing so, they often contradict even correct decisions of the automated system, causing severe decrements in performance (Boskemper et al., 2022). Therefore, cross-checking should be facilitated, allowing people to quickly and reliably judge whether the AI has made the right decision for the right reasons.

A promising way to facilitate cross-checking is to use visualisations that explain why an image classifier has come to a particular decision. Typically, such XAI methods are additional algorithms which are applied on top of the AI classifier and analyse its performance (e.g., Arras et al., 2019; Selvaraju et al., 2017). The XAI method can then overlay images with visual highlights (e.g., heatmaps or bounding boxes) that reflect the weights of the classifier. Such XAI support comes with the promise of calibrating human trust to the actual capabilities of the AI (Leichtmann et al., 2023).

In visual inspection, if the AI has correctly detected a target, the XAI should highlight the target area, and if the AI has correctly rejected the presence of a target, no area should be highlighted. However,

AI systems are error-prone. On the one hand, the AI classifier can err, producing misses or false alarms[1]. For misses, the XAI will probably not highlight any area, and for false alarms, it will highlight an area that does not actually contain a target. In general, such explanations should help people understand the limitations of the AI (Ribeiro et al., 2016), but as we will see in the next section, they also have risks. On the other hand, inadequate highlights may stem from the XAI method itself. When comparing different XAI methods that are supposed to explain one and the same AI classifier, their visual outputs can be strikingly different (Arras et al., 2022; Müller, Thoß, et al., 2023). Thus, even when the AI has made the right decision for the right reasons (e.g., classifying an image as containing a target, based on an actual target) XAI highlights can still be misplaced (sometimes referred to as miscues). Henceforth, we will summarise these different outcomes (i.e., correct detections, correct rejections, misses, false alarms, and misplaced highlights) under the term of "(X)AI accuracy", including problems of both the AI classifier and the XAI method.

Unfortunately, the XAI literature has little to say about the influence of XAI highlights on human visual inspection processes and performance. This is because first, most human-centred evaluations of XAI solely relied on subjective ratings, and second, they were performed on general image material (e.g., images of cats and dogs) instead of focusing on realistic work contexts. However, as XAI highlights are a form of direct cue, inferences about its benefits and costs can be drawn from the literature on cueing in visual inspection tasks. In the following section, we will review key findings from this literature that might provide insights into the potential effects of XAI highlights.

**1.2 Benefits and Costs of Cueing in Visual Inspection Tasks**

The effects of direct target cueing on human visual inspection performance have been investigated for at least three decades (e.g., Alberdi et al., 2004; Chavaillaz et al., 2018; Goh et al., 2005; Hättenschwiler et al., 2018; Huegli et al., 2020; Krupinski et al., 1993a; Maltz & Shinar, 2003; Povyakalo et al., 2004; Warden et al., 2023; Yeh & Wickens, 2001). A common finding is that visual cues generate large benefits when they are accurate, but large costs when they are not. A first aspect of such costs is that people may overly comply with cues, leading to commission errors in case of false alarms (e.g., Maltz & Shinar, 2003; Yeh & Wickens, 2001). A second aspect is that people may overrely on the cues, leading to omission errors in case of misses (e.g., Alberdi et al., 2004). However, most important in the context of this article is a third type of cost that pertains to misplaced highlights: when cues are presented at a wrong location while a non-cued target is located elsewhere, people often fail to identify the target – a phenomenon called attentional tunneling. For instance, in mammography cues on unaffected areas in images that actually did contain cancer led readers to conclude that the patient was cancer-free (Alberdi et al., 2004). Similar findings have been reported in different domains such as radiological image reading (Krupinski et al., 1993a), endoscopic surgery (Dixon et al., 2013), luggage screening (Goh et al., 2005), and military terrain monitoring (Yeh & Wickens, 2001). In all these cases, a more accurate detection of cued targets came at the cost of missing non-cued targets. This suggests that XAI may not only have benefits but costs as well, and that these costs may depend on the specific AI outcome or type of (X)AI accuracy. That is, people may overly comply with XAI highlights in the case of false alarms, overly rely on them in the case of misses, and fail to search for non-cued targets in the case of wrong explanations by misplaced highlights.

---

[1] To avoid confusion, we will reserve these terms from signal detection theory for errors by the AI system, not the human. If we want to differentiate between human error types, we will refer to human misses as "omission errors" and human false alarms as "commission errors".

Several factors affect the type and magnitude of cueing benefits and costs. Perhaps the most important one is cue reliability. Highly reliable direct cues typically improve performance relative to no cues or indirect cues, whereas less reliable cues may fail to generate performance benefits (Chavaillaz et al., 2018; Yeh & Wickens, 2001). At the same time, reliable cues lead to increased reliance and thus wrong decisions when the cue is occasionally wrong (Maltz & Shinar, 2003). A second determinant of cueing effects is the level of unaided performance, resulting from either task difficulty or observer ability. For instance, during airport luggage screening, cues were most helpful when unaided performance was low, either because it was more difficult to discriminate targets from distractors, or because observers were less experienced (Hättenschwiler et al., 2018; Huegli et al., 2020). Similarly, cueing in mammography enhanced the performance of readers with low ability but even deteriorated the performance of readers with high ability (Alberdi et al., 2008). The same cueing paradox that was observed in the case of cue reliability also holds for the level of unaided performance. That is, higher cueing benefits in difficult tasks are accompanied by higher cueing costs of attentional tunneling (Maltz & Shinar, 2003; Povyakalo et al., 2004). A third influence on cueing effects is the visual presentation of the cue. In principle, direct visual cues can take various forms, such as arrows, frames, object outlines, or heatmaps. Again, the most effective cues carry the highest risk of overreliance (Warden et al., 2023). For instance, when tumour areas in radiological images were cued by closed, complete shapes (i.e., full circles, as opposed to dashed-line circles or two vertical bars), people were better at detecting targets inside the cued area but worse at detecting targets outside (Krupinski et al., 1993b).

Taken together, the factors that make cueing most effective also come with the highest risk of overreliance and attentional tunneling. This suggests that any potential benefits of XAI may be accompanied by XAI costs, particularly when the XAI is highly reliable, supports people in difficult tasks, and uses effective visualisations. But can the effects of XAI be readily inferred from the literature on visual cueing? The following section will discuss which limitations of previous research prevent such generalisation and which research gaps need to be addressed.

**1.3 Research Gap and Limitations of Previous Studies**

How does the availability and accuracy of XAI affect the performance and process of visual inspection? As we have shown in the previous section, some aspects of this might be inferred from the past 30 years of cueing studies – only because a visual cue is generated by XAI nowadays, this does not fundamentally change the cognitive mechanisms of using it. It is not so common to directly compare the effects of indirect and direct cues (i.e., black box AI and XAI) as most studies have compared direct cues to manual performance only, but even this has been done before (Chavaillaz et al., 2018; Goh et al., 2005). Thus, why should we need yet another cueing study? In a nutshell, there is a research gap in three aspects, which will be explained in detail in the remainder of this section. Specifically, previous research has only focused on a narrow range of domains, it has seldom differentiated how cueing effects depend on the specific type of cue accuracy, and it has not investigated the attentional mechanisms that underlie these effects of cueing on patterns of behavioural performance.

The first limitation concerns the domains in which the effects of direct visual cueing have been studied. While previous research was conducted in a wide variety of domains (e.g., radiological image reading, luggage screening, military terrain monitoring), these domains still are quite similar on an abstract level in terms of their basic detection requirements. Specifically, all of them are high-risk domains and in all of them it is a central challenge to differentiate targets from distractors. While it certainly is important to understand cueing effects in such domains, it currently in unclear whether and how the results generalise to domains with different characteristics.

A second limitation concerns the independent variable. Cue performance has often been reported as an integrated reliability measure (e.g., accuracy, sensitivity, or positive predictive value). In contrast, it has less often been dissected according to the different ways in which cues can be correct or incorrect (but see Alberdi et al., 2004). However, this would be important to gain a differentiated understanding of XAI effects. Remember that problematic XAI highlights can stem from the AI decision, leading to misses or false alarms, or from the XAI method, leading to misplaced highlights despite correct AI decisions. Thus, it is desirable to understand the effects of XAI on human visual inspection for each type of (X)AI accuracy: correct detections, correct rejections, misses, false alarms, and misplaced highlights.

A third limitation of previous studies concerns the dependent variables. While the cueing literature provides detailed insights about detection performance, we are not aware of any studies that assessed how cueing can influence visual inspection processes. Several studies phrased their results as if they had indeed measured visual attention (e.g., "participants ignored the cue"), but in fact their results do not tell us whether and how participants looked at the cues, targets, or other image areas. What appears to be "ignoring" can also result from opposite tendencies in observers' reactions to cues, which might cancel each other out (Alberdi et al., 2008). The processes that underlie cueing effects can be traced via eye tracking, which is a useful tool to assess trust in automation. For instance, eye tracking can reveal how visual attention changes when an automated system is prone to false alarms versus misses (Wickens et al., 2005), or how observers redistribute their visual attention when the reliability of information sources varies (Lu & Sarter, 2019). In the context of cueing, the only eye tracking studies we are aware of were explicitly not interested in visual search, but only assessed the precision and dispersion of fixations within the small cued area (Krupinski et al., 1993a; Krupinski et al., 1993b). Thus, to date it is unknown how the accuracy of cues affects target search during visual inspection. These limitations of previous research are addressed in the present study.

**1.4 Present Study**

*1.4.1 Contributions of the Present Study*

In the present study, we aimed to address all three research gaps identified in the previous section: the domain specificity of XAI effects, their dependence on (X)AI accuracy, as well as the relation between behavioural performance and the underlying visual inspection processes.

The first factor that distinguishes our study from previous research is the domain in which it is carried out: visual quality control in the food industry. Specifically, participants had to detect faulty bars in images of chocolate moulds. Although this is a realistic task, the visual material is highly structured and uniform (for an example see Figure 1, top left): chocolate bars are spatially arranged in a structure that is the same for each image, and given the standards for product quality, each deviation in the chocolate is a fault. Accordingly, most factors are absent that typically complicate visual inspection. Chocolate moulds do not consist of overlapping translucent image layers, as it is the case for X-ray images (Krupinski et al., 1993a). There is no camouflaging, as it is typical for natural terrain (Maltz & Shinar, 2003; Yeh & Wickens, 2001). There is no visual clutter and objects do not occur in unusual perspectives, as it is the case for luggage screening (Chavaillaz et al., 2018; Goh et al., 2005). Thus, it is not a major challenge to discriminate targets from distractors, as it is the case when observers can easily mistake trees for people (Yeh & Wickens, 2001) or bare explosives for harmless organic mass (Hättenschwiler et al., 2018). This is not to say that our images are easy to work with. However, the challenge will probably be in spotting the fault (i.e., localisation), rather than discriminating it from distractors (i.e., identification). It will be interesting whether typical cueing costs reported in previous research (e.g., attentional tunneling) will generalise to a domain in which more structured and less complex material is to be visually inspected. They might not, given that basic research with simple, artificial stimuli has typically reported cueing benefits only, but no costs. For sure, studying visual inspection with highly complex material is important. However, restricting our focus to only these domains is problematic, because in the future, XAI applications will likely invade all kinds of domains – especially those with high economic impact such as the food industry.

The second contribution of our study is that we ask how the benefits and costs of cueing via XAI depend on the specific outcome of an AI system and its explanation. To this end, we manipulated both XAI availability and (X)AI accuracy in an experimental within-participants design. In half of the experiment, participants worked with black box AI (henceforth called BBAI) and were merely informed about the AI decision before each trial. In the other half, this indirect cue was complemented by a visual explanation in the form of a bounding box (XAI). The accuracy of the (X)AI system was manipulated in five decision outcomes, which appeared at a specific frequency and were characterised by specific highlight positions (see Methods sections for details): correct detections, correct rejections, misses, false alarms, and misplaced highlights. We were particularly interested in the effects of misplaced highlights, where correct AI decisions are wrongly explained. In order to differentiate between these five (X)AI accuracy conditions, a high level of experimental control was mandatory. This is not achievable with an actual AI model and XAI algorithm, both of which can behave in quite unexpected ways (Geirhos et al., 2020; Müller, Dürschmidt, et al., 2023). Therefore, we decided to use a simulated AI system and manually positioned the XAI highlights to generate our stimuli.

The third way in which the present study differed from previous research is that we used eye tracking to elucidate how visual search processes change in response to XAI, particularly in case of wrong explanations. Understanding the relation between eye movements and behavioural patterns of performance is important, because the same performance outcome can result from several underlying processes. Thus, when interpreting performance effects in response to XAI, eye movements can help to differentiate between different causes. For illustration, consider the case of misplaced highlights. If we find that they speed up reactions, this could mean that participants take the highlight for granted without even inspecting it, that they inspect it but refrain from searching the rest of the image, or that they efficiently disengage from the highlighted area, making attentional resources available for searching the actual fault. Conversely, if misplaced highlights do *not* speed up reactions, this could mean that XAI highlights do not affect image processing, or that they focus attention on the highlight at the cost of searching the rest of the image. Similarly, changes in error rates can have different causes. For instance, if participants do not report a fault in case of a misplaced highlight, this could be due to not spotting the fault in the first place, but it could also result from inattentional blindness ("looking but not seeing") or effects of decision-making (concluding that the spotted deviation is not a fault after all) (cf. Manzey et al., 2012). Therefore, we analysed how long participants spent looking at different areas of interest (i.e., fault, highlight, rest of the mould), in how many trials they fixated the fault, and how they differed in their individual strategies of allocating their visual attention across different image areas.

Taken together, we investigated the benefits and costs of XAI-based visual cueing in a new task domain under different (X)AI accuracy conditions, but particularly focused on wrong explanations of correct AI decisions. We did this by examining not only detection performance but also participants' visual search processes as reflected in their eye movements. Thus, readers who do not buy into the idea that XAI leads to fundamental changes in human-technology interaction should feel free to conceive of the present study as simply a cueing study – but one that drills down into the search processes underlying visual inspection performance.

*1.4.2 Hypotheses*

In line with previous reports of cueing benefits, we expected decisions to be faster with XAI than BBAI, because highlights should focus participants' attention on relevant areas. However, we did not expect performance to be more accurate with XAI in general when averaged across (X)AI accuracy conditions, because we assumed that XAI benefits and costs would cancel each other out (cf. Alberdi et al., 2008). Thus, our hypotheses concerned the interaction between XAI availability and (X)AI accuracy. In this section, we will specify these hypotheses about the effects of XAI availability separately for each level of (X)AI accuracy. The following notation will be used: The number of a hypothesis indicates the level of (X)AI accuracy (i.e., H1 refers to hypotheses about correct detections, H2 to hypotheses about correct rejections, etc.). If we expect XAI effects, the letter indicates the respective dependent variable (i.e., the letter a refers to reaction times and the letter b refers to error rates). We do not provide hypotheses for eye movements, because they were analysed in an exploratory manner. If we expect no XAI effects on performance, this is indicated by a zero (e.g., H2-ab0 indicates the absence of XAI effects for correct rejections in reaction times and error rates).

First, let us consider the two (X)AI accuracy conditions with correct outcomes. For *correct detections*, we expected lower reaction times (H1-a) and fewer errors (H1-b) with XAI than BBAI, in line with most previous research on cueing effects (e.g., Chavaillaz et al., 2018; Maltz & Shinar, 2003; Warden et al., 2023; Yeh & Wickens, 2001). Conversely, we did not have a single straightforward hypothesis for *correct rejections* but could imagine outcomes in either of two directions. In this type of trial, no highlights are present even when the system is generally able to provide explanations, and thus XAI trials are perceptually identical to BBAI trials. On the one hand, this could lead to similar performance for XAI and BBAI (H2-ab0). On the other hand, working with an explainable system might generate halo effects (being more trustful towards AI decisions overall), or reassurance effects (considering the absence of cues as strong evidence that everything is okay) (Alberdi et al., 2004). Thus, participants might more readily accept that the mould is intact and thus respond faster (H2-a).

Moreover, there are three (X)AI accuracy conditions with incorrect outcomes. For *misses*, XAI and BBAI again are perceptually identical. Thus, in principle, XAI should not affect performance (H3-ab0). On the other hand, and in line with our reasoning for correct rejections, halo or reassurance effects (Alberdi et al., 2004) might trigger faster responding (H3-a). However, for misses the consequence would be increased error rates (H3-b), in line with the previously reported cueing costs for misses (Alberdi et al., 2004; Goh et al., 2005). For *false alarms*, previous research predicts XAI costs to be particularly severe (Maltz & Shinar, 2003). However, we diverged from this prediction due to the characteristics of our stimuli, specifically the absence of distractors. Thus, when XAI highlights an intact area, we expected faster reactions (H4-a) but no increase in error rates (H4-b0): participants should quickly check the highlighted area, and upon seeing that it is intact, they might correctly conclude that no fault is present, without spending much time searching the rest of the mould.

Finally, the most interesting case is *misplaced highlights*. Here, we expected faster reactions for XAI than BBAI (H5-a), based on the same reasoning as for false alarms: when participants notice that the highlighted area is intact, they might conclude that the chocolate is okay, instead of thoroughly searching the rest of the mould. This strategy should manifest in higher error rates for XAI (H5-b), in line with previous findings (Goh et al., 2005; Yeh & Wickens, 2001), because participants would overlook the fault located elsewhere. However, note that for misplaced highlights, performance data alone are not interpretable. This is because reactions would be both fast and accurate if participants blindly relied on the AI classifier, simply agreeing when it reports a fault, without even checking the highlight. We do not consider this particularly likely, because when people are given the chance to cross-check, they usually do so (Manzey et al., 2014; Müller et al., 2020; Müller, Mangold, et al., 2023). Still, to exclude this option, it is mandatory to trace the search processes underlying participants' responses. To better understand the origins of performance effects, we report eye movement results both for the entire sample and individual participants.

To test our hypotheses, we conducted two experiments. In Experiment 1, we gained first insights into the visual inspection processes during chocolate quality control. As this was the first study in our lab using the chocolate scenario, we included a broad range of different fault types. In Experiment 2, we replicated the task with more controlled stimulus material. All stimuli, data, and syntax files of this study are made available via the Open Science Framework: https://osf.io/sxqmn/

## 2 Experiment 1

### 2.1 Method

*2.1.1 Participants*

Thirty members of the TUD Dresden University of Technology participant pool took part in the experiment in exchange for course credit or a payment of 8 € per hour. The sample consisted of 22 female and 8 male participants with an age range of 18 to 43 years ($M$ = 25.1, $SD$ = 5.8). Only participants with normal vision were included. All procedures followed the principles of the Declaration of Helsinki.

*2.1.2 Apparatus and Stimuli*

*Lab setup*. The experiment took place in a lab room at TUD. Stimuli were presented on a 24" LCD display with a resolution of 1920 by 1080 pixels at a refresh rate of 60 Hz. A standard QWERTZ computer keyboard was used for keypress responses. Eye movements were tracked monocularly at 1000 Hz using the EyeLink 1000 infrared eye tracking system (SR Research Ltd., Ontario, Canada) with a chin rest and a viewing distance of 93 cm. This system allows for an online detection of saccades and fixations, and a spatial accuracy < 0.5 degrees of visual angle. To identify fixations and saccades, we used the saccade detection algorithm supplied by SR Research. This algorithm detects saccades via deflections in gaze position > 0.1 deg, with a minimum velocity of 30 deg/sec and a minimum acceleration of 8000 deg/sec$^2$, maintained for at least 4 ms. A standard nine-point calibration was used at the start of each block, and repeated during the experiment in case a loss of accuracy was detected during drift correction before a given trial.

*Stimuli used during the experiment*. The stimulus material is illustrated in Figure 1. All stimuli were shown on a black background, and all text was presented in German. There were two main types of stimuli: AI decision screens and search screens. The AI decisions and XAI highlights presented in those screens were generated manually (i.e., we used a simulated instead of a real AI system). AI decision screens presented an indirect cue as a text message (Calibri, 50 pt), indicating whether the chocolate mould was classified as faulty or not. Thus, it either read "Machine Learning: chocolate is faulty!" (in red font) or "Machine Learning: chocolate is okay!" (in white font). Search screens (see Figure 1, top left) consisted of a centrally positioned chocolate mould with a size of 1180 x 615 pixels (21.0 x 11.1 deg). The mould contained 81 chocolate bars aligned in three rows, with each bar subtending an area of 34 x 144 pixels (0.6 x 2.6 deg). The mould was a natural photograph, thus different areas varied in their lighting and shading, and slight irregularities in the chocolate were present. However, the same mould was used as a background image in each trial, so participants could easily learn not to mistake repeating irregularities for faults.

Each search screen (mould) could either contain one fault and one highlight, either of them, or neither of them. Depending on (X)AI accuracy, different relations between these faults and highlights were possible (see Figure 1, right). There either was a fault and a highlight at the same location (i.e., correct detections), a fault and a highlight at a different location (i.e., misplaced highlights), a fault but no highlight (i.e., misses), no fault but a highlight (i.e., false alarms), or no fault and no highlight (i.e., correct rejections).

When the chocolate actually was faulty, one bar in the mould had a small defect of about 20 x 20 pixels (0.4 x 0.4 deg). Six types of faults were used (see Figure 1, bottom left): air bubbles, fractures, hair, imprints, nuts, and scratches. Each fault type was presented equally often, and in three versions that differed in their rotational orientation and image details. During fault-present trials, exactly one fault was superimposed on the background mould image. The faulty chocolate bar was determined pseudo-randomly, with the following restrictions: the first and last bar in a row never contained the fault, and each of the remaining bars contained it about equally often (i.e., 3-4 times during the entire experiment). To compute the fault position on an individual chocolate bar, faults were horizontally centred on the bar and vertically shifted by a random value, so each vertical bar position was about equally likely to contain the fault. Fault positions were computed for all (X)AI accuracy conditions, even those with non-faulty chocolate (i.e., correct rejections and false alarms, in these cases the faults were rendered invisible). This was done to determine baseline probabilities of fixating the respective areas when no fault was actually present.

Our simulated XAI highlights were visualised as bounding boxes with an area of 80 x 80 pixels (1.4 x 1.4 deg) and a red outline. Bounding boxes are one of many possible visualisations for XAI highlights (e.g., Nourani et al., 2019), and although they are less common than heatmaps, they come with the advantage of not occluding the target. In contrast to other visualisations like outlines of the most relevant area (e.g., Sundararajan et al., 2019), their size and shape can be held constant across different fault types. Moreover, cue shapes that are more complete and even are more effective than those with less complete boundaries (Krupinski et al., 1993b).

When the highlight targeted the actual fault (i.e., correct detections), it was centred on the fault position. When it did not (i.e., misplaced highlights), it was randomly placed on one of the remaining bars, with the exception that it never appeared adjacent to the faulty bar (i.e., three bars to the left and right in same row, two bars to the left and right in rows above and below). Moreover, highlights never appeared on the first or last bar in a row. When a highlight but no fault was present (i.e., false alarms), the highlight was centred on the invisible fault position. To determine baseline probabilities of fixating a highlight, highlight positions were also computed but rendered invisible for all conditions without highlights (i.e., all BBAI trials, correct rejections and misses for XAI trials).

Aside from the main stimuli, three other screens were used. First, at the start of the experiment participants saw a screen on which they had to input their demographic data (i.e., age and gender). Second, before each block, instruction screens summarised the respective task. These screen informed participants whether AI decisions and XAI highlights would be available in the following block, and reminded them of the correct response key mapping. Third, a time-out screen informed participants when they had exceeded the available time limit (i.e., "Too slow!" in red font).

*Additional materials*. Instructions were provided in a Powerpoint presentation. It informed participants that their task was to monitor chocolate moulds for faults, and either discard the mould when there is a fault, or proceed to the next mould as quickly as possible when three is none. An example of an intact mould was shown, as well as an enlarged version of the six fault types. Moreover, participants learned that they would be supported by AI in some parts of the experiment. The trial procedure was outlined with examples of the respective screens to explain the difference between XAI and BBAI. Participants did not receive information about (X)AI accuracy, and neither were any strategies suggested to them.

**Figure 1**

Stimuli with search screen (top left), faults (bottom left) and (X)AI accuracy conditions (right). In the right-hand part of the figure, black and white boxes represent simplified versions of the AI decision screen and search screen, respectively. In the latter, brown dots represent a fault and red boxes represent XAI highlights. The relations between these faults and highlights are illustrated for all five levels of (X)AI accuracy. A highlight can either be positioned at the location of the fault (for correct detections) or at a location that is not faulty (for misplaced highlights and false alarms).

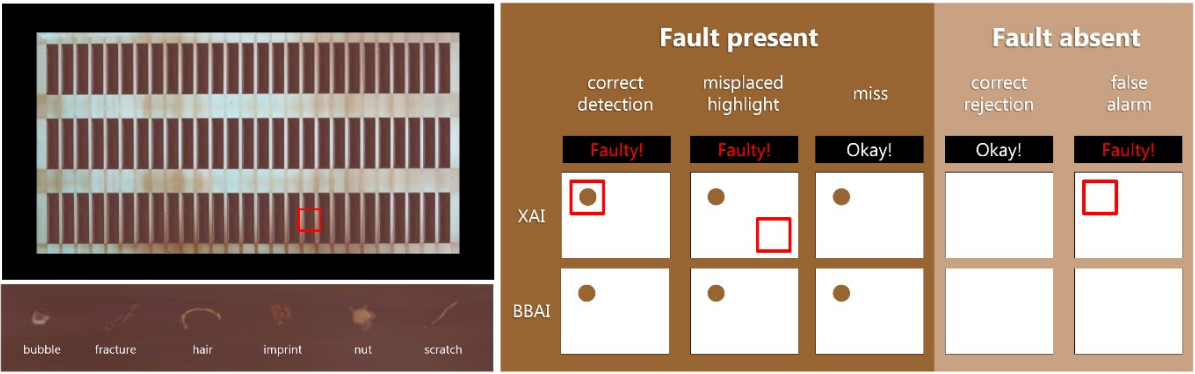

2.1.3 Procedure

An overview of the procedure is provided in Figure 2. Experimental sessions took about 60 minutes. Upon entering the lab, participants were informed about the overall study procedure, provided written informed consent, and went through the instruction. After this, they started the experiment. The basic trial procedure was as follows: a trial started with a drift correction and by pressing the Space bar while fixating it, participants proceeded to the AI decision screen. This screen was presented for 1500 ms and was then replaced by the search screen. Participants had to submit their decision about whether the chocolate was faulty or okay by pressing a right (X) or left (N) key, respectively. If no response was submitted within 5000 ms, the time-out message appeared for 1000 ms and the trial was terminated.

Moreover, while viewing the search screen, participants had the opportunity to view the intact mould: pressing the Space bar first led to the presentation of a blank screen for 500 ms, and then revealed an image of the mould with no fault. Pressing the Space bar again led to another blank screen for 500 ms and then brought participants back to the search screen.

The experiment consisted of six blocks: one practice block, three blocks of unaided performance, and two AI blocks (one with BBAI, one with XAI). During practice blocks, participants' task was to indicate whether the chocolate was faulty or not, after seeing an AI decision before each trial but no XAI highlights (i.e., BBAI). Ten practice trials appeared in random order, with seven correct detections and three correct rejections. The three unaided blocks were placed before, between, and after the two AI blocks. They had been added to investigate a research question within a student project conducted by the second and third author, but we do not report the respective analyses as they are irrelevant for the purpose of this article. In unaided blocks, participants' task was to indicate whether the chocolate was faulty or not, but no BBAI or XAI support was available. Unaided blocks consisted of 30 trials, with 20 of them containing a fault.

The two AI blocks varied whether XAI support was available or not (XAI and BBAI, respectively), and block order was counterbalanced between participants. In BBAI blocks, participants received an indirect cue about the AI decision before seeing the mould, whereas in XAI blocks, they additionally received a direct cue in the form of an XAI highlight whenever the AI had classified a mould as faulty. To assess how XAI benefits and costs depended on (X)AI accuracy, we needed comparable and thus completely parallel blocks. Therefore, each (X)AI accuracy condition in the XAI block of even-numbered participants had its respective twin trials in the BBAI block of odd-numbered participants. The respective trials only differed in whether a highlight was visible, but were otherwise identical, for instance in terms of their numbers, positions in the block, or fault locations. Therefore, we coded all (X)AI accuracy conditions for both block types, including misplaced highlights, although no highlights were shown in BBAI blocks, and thus, the trials coded as "misplaced highlight" were in fact correct detections (the same approach was used by Alberdi et al., 2004).

An overview of the trial numbers and characteristics of all (X)AI accuracy conditions is provided in Table 1. Each AI block consisted of 90 trials, with two thirds being fault-present trials, corresponding to a signal rate of 66 %. AI reliability and cue reliability had the same proportions: 70 % of the AI decisions and 70 % of the XAI highlights were correct. The correct AI decisions consisted of two thirds of correct detections (42 trials) and one third of correct rejections (21 trials). For *correct detections*, the AI decision screen stated that the AI had classified the chocolate as faulty, the search screen actually contained a fault, and in XAI blocks this fault was correctly highlighted. For *correct rejections*, the AI decision screen stated that the AI had classified the chocolate as okay, and the search screen indeed neither contained a fault nor a highlight (even in XAI blocks). The incorrect AI decisions consisted of one third of misses, false alarms, and misplaced highlights, respectively (9 trials each). For *misses*, the AI classified the chocolate as okay, but the search screen contained a non-highlighted fault. For *false alarms*, the AI classified the chocolate as faulty and a highlight was shown in XAI blocks, but the highlighted area was intact and no other fault present. For *misplaced highlights*, the AI also classified the chocolate as faulty while the highlighted area was intact. However, a non-highlighted fault was located in another area of the mould. Note that in BBAI blocks, this trial type was perceptually identical to correct detections. The order of trials from different (X)AI accuracy conditions within a block was randomised individually for each participant. To avoid confounds with stimulus difficulty, we also made sure that there were no systematic differences in the assignment of the six fault types (e.g., air bubbles, fractures) to the five (X)AI accuracy conditions, and that this assignment was identical for XAI and BBAI.

**Figure 2**

Overview of the procedure with experimental blocks (left) and events within a trial (right)

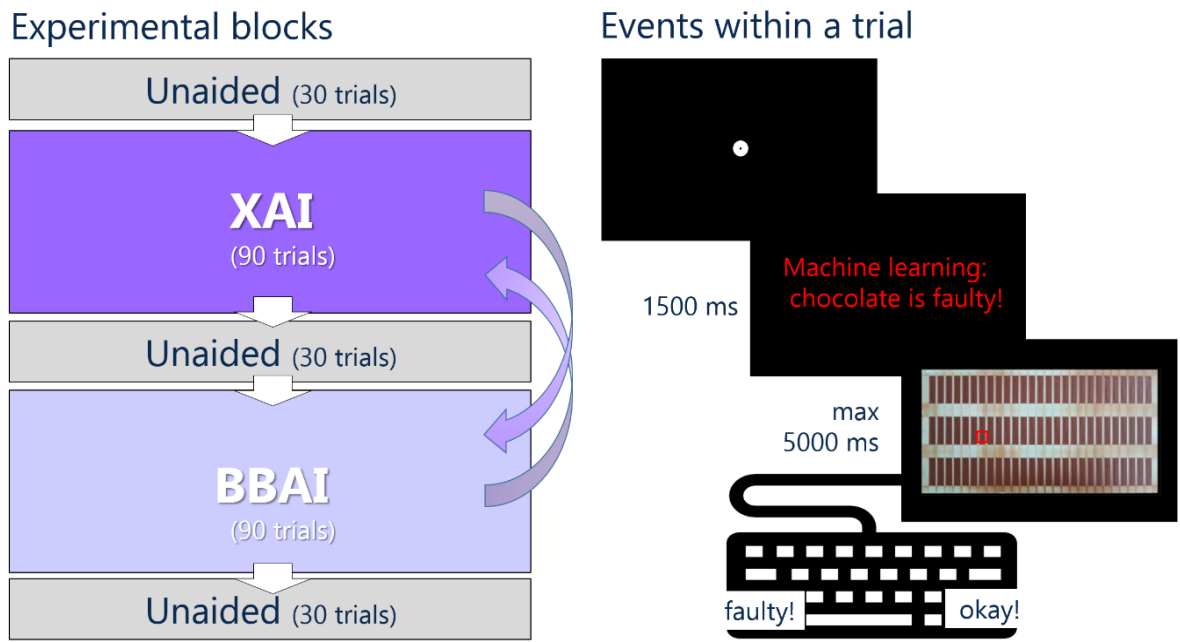

**Table 1**

Characterisations and numbers trials within an AI block, arranged by (X)AI accuracy conditions. As the same trial numbers were used in XAI and BBAI blocks, we only present them once.

|  | Correct detection | Misplaced highlight | Miss | Correct rejection | False alarm | Total per AI block |
|---|---|---|---|---|---|---|
| Fault presence | Present | Present | Present | Absent | Absent |  |
| AI decision | Faulty | Faulty | Okay | Okay | Faulty |  |
| (X)AI correctness | Correct | Incorrect | Incorrect | Correct | Incorrect |  |
| Experiment 1 | 42 | 9 | 9 | 21 | 9 | 90 |
| Experiment 2 | 70 | 15 | 15 | 35 | 15 | 150 |
| Percent | 46.7 | 10 | 10 | 23.3 | 10 | 100 |

*2.1.4 Data Analysis*

We analysed performance and eye movement data in the AI blocks. Performance was assessed via mean reaction times and error rates. Reaction time was defined as the latency from the onset of the search screen (i.e., chocolate mould) until participants pressed a key to indicate whether the chocolate was faulty or not. Error rates were defined as the percentage of trials in which participants made the wrong decision, either judging the chocolate as okay while it actually was faulty, or vice versa. We eliminated trials in which no decision was submitted, because the deadline was exceeded (.63 % of the data). For the remaining data, mean values were computed (see Table 2) and compared between the experimental conditions.

Eye movements were analysed in an exploratory manner to gain insights into participants' visual inspection processes, particularly in misplaced highlight trials. We performed a location-based analysis of eye movements, determining whether and for how long participants fixated three different areas of interest (AOIs): the fault, the highlight, and the rest of the mould. To this end, we first determined for each fixation (detected by the eye tracing software as described above) in which AOI it had landed, based on its x and y coordinates. AOI size of the fault and highlight AOI was set to 100 x 100 px (1.8 x 1.8 deg) for all analyses. We considered only fixations on the search screen, excluded the first fixation (i.e., the one that started on the AI decision screen), but included the last fixation (i.e., the one that was interrupted by the end of the trial). These AOI-flagged fixations were then analysed in three ways. First, we computed the average dwell times per trial (i.e., sum of all fixation durations) for each of the three AOIs (i.e., fault, highlight, rest of the mould). Second, we computed the percentage of trials in which participants fixated the fault AOI. Third, we conducted two exploratory analyses of individual participants' search strategies, which will be explained in the respective part of the Results section. Due to technical problems, one eye movement file was defective, thus only the data of 29 participants entered the eye movement analyses.

For statistical analyses, we conducted paired t-tests to check for general XAI effects (pooled across all trials) and 5 (*(X)AI accuracy: correct detection, misplaced highlight, miss, correct rejection, false alarm*) x 2 (*XAI availability: XAI, BBAI*) repeated measures ANOVAs to check how these XAI effects depended on (X)AI accuracy. An alpha value of $p = .05$ was used to determine statistical significance, and all pairwise comparisons were performed with Bonferroni correction. If the sphericity assumption was violated, a Greenhouse-Geisser correction was applied and the degrees of freedom were adjusted accordingly. Moreover, to compare dwell dimes in specific AOIs between XAI and BBAI, we performed paired t-tests. We conducted correlation analysis (Pearson) to determine the relation between participants' reaction times and error rates. Other analyses are described in the respective parts of the Results section.

## 2.2 Results

*2.2.1 Performance*

In the following sections, we will include (X)AI accuracy as a factor in our statistical analyses, so that each condition enters the analysis with equal weight. However, to get a realistic estimate of overall XAI effects, it needs to be considered that correct (X)AI (i.e., correct detections, correct rejections) was presented more frequently than incorrect (X)AI (i.e., misplaced highlights, misses, and false alarms). Therefore, before turning to our main results, we will report overall XAI effects by pooling all trials of the AI blocks, so that all (X)AI accuracy conditions enter with a weight that corresponds to their trial numbers. This yielded a significant XAI benefit for reaction times, $t(29) = 8.598$, $p < .001$, $d = 1.570$, indicating that participants responded faster with XAI than BBAI (1.6 vs. 1.9 sec, respectively). There also was a significant XAI benefit for error rates, $t(29) = 6.382$, $p < .001$, $d = 1.165$, reflecting fewer errors with XAI than BBAI (8.7 vs. 14.3 %, respectively).

*Reaction times*. The ANOVA revealed a main effect of (X)AI accuracy, $F(1.3,38.1) = 87.499$, $p < .001$, $\eta p^2 = .751$, a main effect of XAI availability, $F(1,29) = 30.622$, $p < .001$, $\eta p^2 = .514$, and an interaction of both factors, $F(3.1,89.6) = 24.326$, $p < .001$, $\eta p^2 = .456$ (see Figure 3A). Responses were fastest for correct detections (1.3 sec), slowest when faults were absent (2.4 and 2.6 for correct rejections and false alarms, respectively), and intermediate when faults were present but not identified by the AI (1.6 sec for misses) or wrongly explained by the XAI (1.6 sec for misplaced highlights). All pairwise comparisons were significant, all $p$s < .001, except for the comparison between misplaced highlights and misses, $p > .9$. Participants responded faster with XAI than BBAI (1.9 vs. 2.1 sec, respectively), but the interaction revealed that this XAI benefit was restricted to correct detections, correct rejections, and false alarms, all $p$s < .001. Not surprisingly, a large XAI benefit (540 ms) was observed for correct detections, but a somewhat smaller XAI benefit (285 ms) was also present for correct rejections. Moreover, there was a large XAI benefit (538 ms) for false alarms. Misplaced highlights showed a non-significant trend for XAI costs (133 ms), $p = .088$, and no XAI effects were found for misses, $p = .221$.

*Error rates*. The ANOVA revealed a main effect of (X)AI accuracy, $F(2.3,65.7) = 18.523$, $p < .001$, $\eta p^2 = .390$. The main effect of XAI availability was not significant, $F(1,29) = .173$, $p = .681$, $\eta p^2 = .006$, but XAI availability interacted with (X)AI accuracy, $F(2.6,74.6) = 11.042$, $p < .001$, $\eta p^2 = .276$ (see Figure 3B). The main effect of (X)AI accuracy indicated that error rates were highest when faults were present but not correctly indicated (22.3 and 24.6 % for misplaced highlights and misses, respectively), low when faults were absent (1.5 and 8.0 % for correct rejections and false alarms, respectively), and intermediate for correct detections (12.1 %). Most pairwise comparisons between these three types of trials were significant, all $p$s < .042, (except for the difference between correct detections and false alarms, $p > .1$), while the comparisons within them were not, both $p$s > .282. The absence of a main effect of XAI availability indicated that XAI did not improve performance overall, compared to BBAI (13.5 vs. 13.9 %, respectively). Instead, the presence and direction of XAI effects strongly depended on (X)AI accuracy. While XAI reduced errors by 14.3 % for correct detections, $p < .001$, it increased errors by 10.2 % for misplaced highlights, $p = .012$. For the remaining (X)AI accuracy conditions, differences between XAI and BBAI were not significant, all $p$s > .202.

**Figure 3**

Performance results for Experiment 1, depending on XAI availability and (X)AI accuracy. (A) Mean reaction times and (B) error rates. Error bars represent standard errors of the mean.

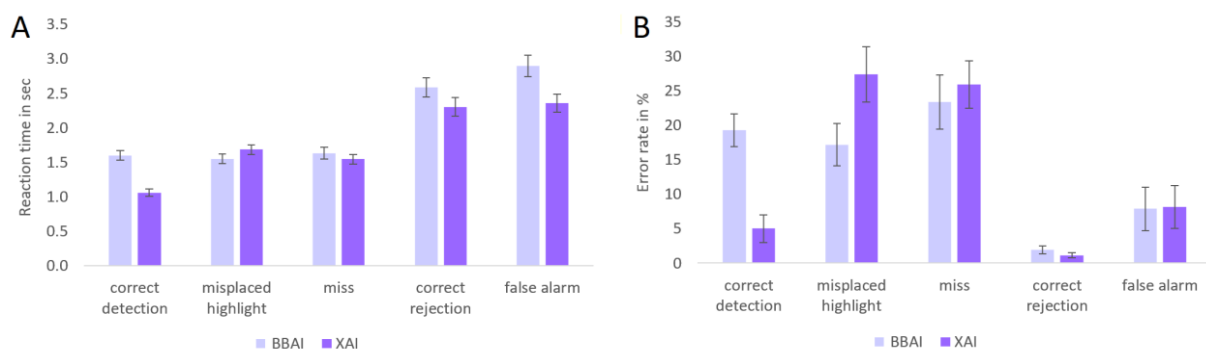

**Table 2**

Mean reaction times and error rates (with standard deviations in parentheses) for both experiments, depending on XAI availability and (X)AI accuracy. Exp = experiment, RT = reaction time, PE = percent errors. Positive difference values between BBAI and XAI reflect XAI benefits and negative values reflect XAI costs; significant differences are presented in bold. * = $p < .05$, ** = $p < .01$, *** = $p < .001$, n.s. = not significant.

|      |         |            | Correct detection | Misplaced highlight | Miss | Correct rejection | False alarm |
|------|---------|------------|-------------------|---------------------|------|-------------------|-------------|
| Exp1 | RT (sec) | BBAI      | 1.6 (.4)          | 1.6 (.4)            | 1.6 (.5) | 2.6 (.8)      | 2.9 (.8)    |
|      |         | XAI        | 1.1 (.3)          | 1.7 (.4)            | 1.5 (.4) | 2.3 (.7)      | 2.4 (.7)    |
|      |         | Difference | **.5 (\*\*\*)**   | -.1 (n.s.)          | .1 (n.s.) | **.3 (\*\*\*)** | **.5 (\*\*\*)** |
|      | PE (%)  | BBAI       | 19.3 (13.1)       | 17.2 (16.7)         | 23.3 (21.5) | 1.9 (3.3)   | 7.9 (17.3)  |
|      |         | XAI        | 5.0 (10.9)        | 27.4 (22.0)         | 25.9 (18.8) | 1.1 (2.1)   | 8.1 (17.0)  |
|      |         | Difference | **14.3 (\*\*\*)** | **-10.2 (\*)**      | -2.6 (n.s.) | .8 (n.s.)   | -.2 (n.s.)  |
| Exp2 | RT (sec) | BBAI      | 3.2 (.4)          | 3.1 (.5)            | 3.2 (.5) | 3.7 (.3)      | 4.1 (.4)    |
|      |         | XAI        | 1.2 (.3)          | 2.9 (.8)            | 3.0 (.5) | 3.5 (.6)      | 3.1 (1.0)   |
|      |         | Difference | **2.0 (\*\*\*)**  | .2 (n.s.)           | .2 (n.s.) | **.2 (\*\*)** | **1.0 (\*\*\*)** |
|      | PE (%)  | BBAI       | 32.7 (15.0)       | 29.3 (18.4)         | 43.1 (20.7) | 2.8 (10.4)  | 9.5 (17.3)  |
|      |         | XAI        | 1.9 (4.1)         | 54.1 (27.4)         | 50.9 (20.0) | 2.2 (5.3)   | 6.7 (10.9)  |
|      |         | Difference | **30.8 (\*\*\*)** | **-24.8 (\*\*\*)**  | -7.8 (n.s.) | .6 (n.s.)   | 2.8 (n.s.)  |

*Stimulus difficulty*. An exploratory analysis of stimulus difficulty showed large variations in performance between the six fault types. Mean reaction times were lowest for bubbles and nuts (1.1 sec for both), intermediate for hair, imprints, and scratches (1.6, 1.8 and 1.8 sec, respectively), and highest for fractures (2.2 sec). These results were mirrored in the error rates, with the fewest errors for bubbles and nuts (4.1 % for both), medium error rates for hair, imprints, and scratches (18.5, 31.9, and 29.3 %, respectively), and a particularly high error rate for fractures (45.3 %).

2.2.2 Eye Movements

*Dwell times within different AOIs*. Contrary to our hypotheses, XAI did not speed up reactions for misplaced highlights, but we even observed a non-significant XAI cost. Does this mean that participants did search the mould thoroughly, despite the wrong explanation? To investigate what they actually looked at before responding, we decomposed the total reaction times into dwell times within each AOI (i.e., sum of all fixation durations on the fault, highlight, or rest of the mould). A t-test indicated that for misplaced highlights, participants spent less time searching the rest of the mould with XAI than BBAI (.7 vs. 1.2 sec), $t(28) = -4.914$, $p < .001$, $d = .912$. The same was true for false alarms, with less time on the rest of the mould with XAI than BBAI (1.5 vs. 2.7 sec), $t(28) = -8.942$, $p < .001$, $d = 1.661$. Dwell times in the other AOIs (i.e., fault and highlight) were similar across conditions (see Figure 4A).

*Fixations within the fault AOI*. XAI led to increased error rates for misplaced highlights, and the dwell time analysis suggested that participants did not thoroughly scan the rest of the mould. Thus, it seems likely that they did not spot the fault. However, failures to report targets can have other reasons, such as "looking but not seeing", or biases in decision-making and reporting (Krupinski et al., 1993a).

Therefore, we computed the percentage of trials in which participants fixated the fault AOI, and compared it between XAI and BBAI for all (X)AI accuracy conditions in which faults were present. For statistical comparison, we performed a 3 (*(X)AI accuracy: correct detection, misplaced highlight, miss*) x 2 (*XAI availability: XAI, BBAI*) repeated measures ANOVA. This analysis yielded a main effect of (X)AI accuracy, $F(1.5, 42.7) = 14.128$, $p < .001$, $\eta_p^2 = .335$, and while XAI availability did not produce a main effect, $F(1, 28) = .062$, $p = .806$, $\eta_p^2 = .002$, it significantly interacted with (X)AI accuracy, $F(1.6, 45.3) = 18.713$, $p < .001$, $\eta_p^2 = .401$. The main effect of (X)AI accuracy indicated that the fault AOI was fixated in more trials for correct detections than misplaced highlights and misses (92.0 vs. 80.7 and 83.9 %, respectively), both $p$s $< .001$, while the latter two conditions did not differ, $p = .537$. For correct detections, participants fixated the fault more often with XAI than BBAI (98.4 vs. 85.6 %), $p < .001$. The opposite was found for misplaced highlights, where XAI reduced the percentage of fault fixations (75.5 vs. 85.8 %), $p = .046$. For misses, there was no difference between XAI and BBAI (83.9 and 83.9), $p > .9$.

*Individual strategies*. The significant XAI effects for misplaced highlights reported so far do not tell us whether reductions in search effort are indicative of a general behavioural tendency adopted by the majority of participants, or merely due to a subset of participants with extreme responses. Thus, it seems worthwhile to take a closer look at interindividual differences. We did this in two steps.

In the first step, we classified each misplaced highlight trial of each participant according to its decision and the evidence on which this decision was based (see Figure 4C). This led to four classes of trials. When participants had made the (correct) decision that the chocolate was faulty, they could either have fixated the fault (i.e., justified "faulty") or not fixated it (i.e., non-justified "faulty"). Similarly, when participants made the (incorrect) decision that the chocolate was okay, there were trials in which they had actually searched the rest of the mould after fixating the empty highlight (i.e., justified "okay") but also trials in which they had not (i.e., non-justified "okay"). Figure 4C suggests that decisions with a complete lack of evidence (dark green and dark red sections) were rare – most participants actually did look at the mould and/or fault in the majority of trials.

However, this alone does not tell us how thoroughly they searched, as even a single fixation could classify a trial as "justified". Therefore, in the second step we plotted the duration of search (see Figure 4D): the average time per trial a participant spent fixating the rest of the mould, split by whether the trial was subsequently judged as okay or faulty. The plateau in the green bars in the Figure 4D suggests that that most participants spent a medium time searching, while only a few participants cut their search very short and a few searched extensively. Still, the figure suggests a rather gradual variation, instead of a clear distinction between discrete strategies.

**Figure 4**

Eye movement results for Experiment 1. (A) Dwell times within different AOIs, (B) Trials with fixations within the fault AOI. The data for correct rejections and false alarms appear in grey as no faults were visible. However, we still plotted the data as a baseline for visual comparison: bars for correct rejections and the lower bar for false alarms represent the likelihood of accidentally landing in an AOI with the size and location of the fault AOI, while the higher bar represents the likelihood of fixating an empty highlight. Error bars represent standard errors of the mean. The last two parts of the figure depict individual participants' search behaviour for misplaced highlights: (C) Trialwise combination of decision and evidence, (D) Time spent searching the rest of the mould

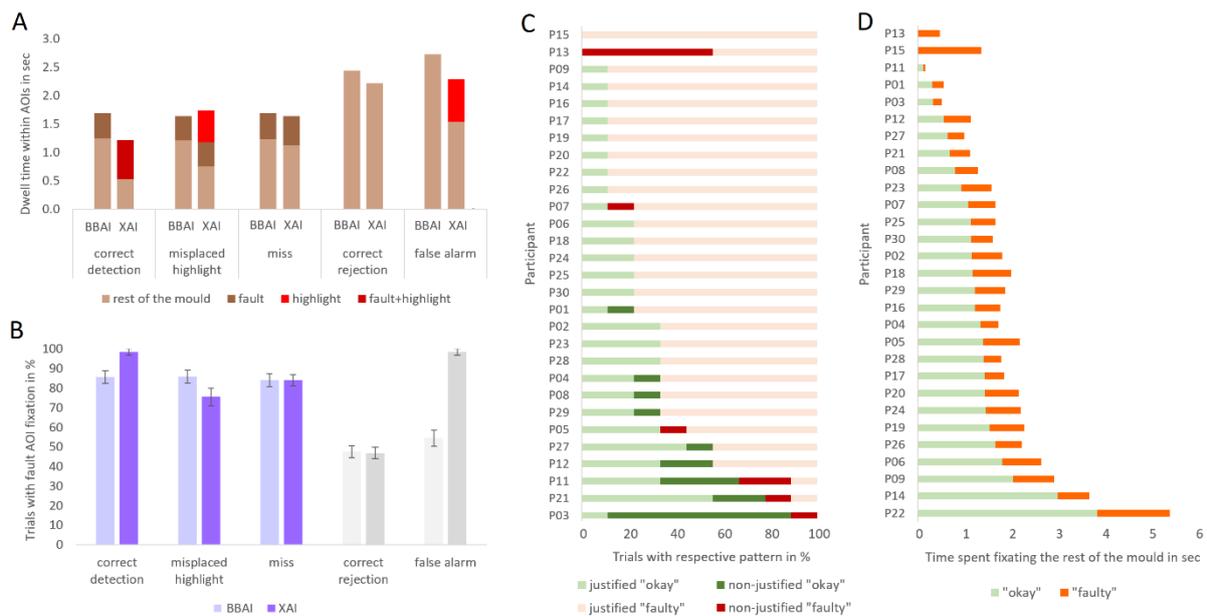

## 2.3. Discussion

Experiment 1 showed that the effects of XAI highlights strongly depend on (X)AI accuracy. For correct detections, we found the expected XAI benefits, both in terms of faster responses and fewer errors (supporting hypotheses H1-a and H1-b). We also found XAI reaction time benefits for correct rejections, although no highlights were shown (supporting H2-a, but contradicting H2-ab0). An explanation could be that when participants knew the AI system was able to generate highlights, the absence of such highlights increased their confidence in the classification of the chocolate being intact. This resembles the conclusion from a mammography study, where cueing benefits also occurred for (and were restricted to) correct rejections (Alberdi et al., 2004). Still, it is not trivial to find XAI benefits when no XAI is shown, so this finding should be replicated before drawing conclusions. So far, our results replicated previous findings of cueing benefits when the cueing is correct.

The picture gets a bit more diverse for the effects of incorrect (X)AI. First, for misses we neither found XAI benefits for reaction times, nor XAI costs for errors. This contradicts our hypotheses H3-a and H3-b, and seems at odds with what we just argued for correct rejections: if the absence of highlights when using a system with XAI capabilities had reassured participants that there is no fault, should not this have resulted in omission errors for misses? However, we think we should replicate this surprising result before drawing conclusions about the underlying mechanisms. Second, for false alarms participants were considerably faster with XAI, while their accuracy was not affected (in line with H4-a and H4-b0). This suggests that they did check the highlighted area, saw that it was intact, and thus concluded that the entire mould was intact. Although this is not a very thorough strategy, it would lead to fast and accurate performance. However, the same strategy is problematic for misplaced highlights, because it can lead participants to terminate their search prematurely and submit an incorrect response (as predicted by H5-b). But did participants actually do this? At first glance, this seems to be contradicted by the fact that XAI did not reduce reaction times (refuting H5-a). However, this can be explained by looking at the dwell times: with XAI, participants spent quite some time fixating the empty highlight, but less time searching the rest of the mould. This led to the absence of a net effect in reaction times. Another factor contributing to the absence of reaction time differences is the large interindividual variation. In fact, most participants did search the rest of the mould before concluding that the chocolate was okay, and their search times varied rather gradually.

Although Experiment 1 produced some interesting findings, two methodological problems complicate their interpretation. First, we observed large variations in stimulus difficulty. These were due to the non-uniform lighting across the mould, as well as the visual appearance of different fault types (e.g., bubbles were much easier to see than fractures). The ease of spotting a fault can be assumed to have major effects on the performance and process of visual inspection – if the fault pops out immediately, neither will omission errors occur, nor will any search activity be needed. The present results might well be a mixture of different phenomena, such as omission errors despite thorough scanning for difficult faults, combined with accurate performance despite a complete lack of scanning for pop-out faults. A second methodological problem is the presence of unaided blocks before, between, and after the AI blocks. These blocks were not relevant for the purposes of the present article, but required participants to look for faults on their own. It is possible that this activity affected their visual search strategies in subsequent AI blocks, for instance by fostering a higher level of self-reliance. Moreover, it only allowed us to use nine instances of each problematic (X)AI accuracy condition to keep the experiment at an acceptable length. Combined with the stimulus-induced variance, this puts the results on shaky ground, which is why we conducted Experiment 2.

## 3 Experiment 2

The aim of Experiment 2 was to replicate the results of Experiment 1 while eliminating its methodological problems. To achieve a uniform, high level of stimulus difficulty, we changed the background image of the chocolate mould into one with invariant lighting, and only used the most difficult fault type. We also eliminated the unaided blocks and instead increased the trial numbers in the AI blocks. Moreover, we removed the option to check the image of the intact mould. Given these methodological changes, our main question was whether Experiment 2 would support the conclusions from Experiment 1 for misplaced highlight trials: that with XAI, participants focus on checking whether the highlight is accurate, and discount the AI decision when it is not, without thoroughly searching for other aspects of the data that might have led to this decision. At the same time, we expected to find large interindividual differences again, with a gradual variation between thoroughly searching the entire mould versus only focusing on the XAI highlight.

### 3.1 Method

*3.1.1 Participants*

Thirty members of the TUD Dresden University of Technology participant pool took part in the study in exchange for course credit or a payment of 8 € per hour. None of them had participated in Experiment 1 or any other chocolate experiments from our lab. Four participants were excluded from the analyses and replaced by new participants. One of them misunderstood the task (i.e., evaluated the AI instead of the chocolate), one fell asleep during the experiment, and two were unable to see the faults and thus made less than 1 % correct decisions in fault-present trials of the BBAI block. The final sample consisted of 20 female and 10 male participants with an age range of 18 to 45 years ($M$ = 25.3, $SD$ = 5.8). Only participants with normal vision were included. All procedures followed the principles of the Declaration of Helsinki.

*3.1.2 Apparatus and stimuli*

The apparatus was identical to that used in Experiment 1. The stimulus material was similar, with the following changes. First, a uniform lighting of the chocolate mould was achieved by copying one particular chocolate bar and pasting it to all other 80 positions of the mould. Second, we only used fractures, which were the most difficult fault type. We retained the three versions of this fault, thus fractures appeared with different rotational orientations. Third, the Powerpoint instruction was adjusted accordingly, removing all information about different fault types and unaided blocks.

*3.1.3 Procedure*

The overall procedure was similar to that of Experiment 1, with the following changes. First, we removed the unaided blocks. Second, we instead increased the trial numbers in the AI blocks from 90 to 150, so that each problematic (X)AI accuracy condition appeared 15 times (see Table 1). Third, participants no longer had the opportunity to view the intact mould during a trial.

*3.1.4 Data analysis*

The data analysis procedure was identical to that of Experiment 1. We excluded all trials that exceeded the deadline of 5000 ms (4.04 % of the data).

**3.2 Results**

Again, we first assessed overall XAI effects by pooling across all trials, so that (X)AI accuracy conditions with higher trials numbers had more weight. There was a significant XAI benefit for reaction times, $t(29) = 21.172$, $p < .001$, $d = 3.865$, indicating that participants responded faster with XAI than BBAI (2.2 vs. 3.4 sec, respectively). There also was an XAI benefit for error rates, $t(29) = 7.969$, $p < .001$, $d = 1.455$, which resulted from participants making fewer errors with XAI than BBAI (12.5 vs. 24.3 %, respectively).

*3.2.1 Performance*

*Reaction times.* The ANOVA revealed a main effect of (X)AI accuracy, $F(2.0, 59.2) = 119.396$, $p < .001$, $\eta p^2 = .805$, a main effect of XAI availability, $F(1,29) = 81.269$, $p < .001$, $\eta p^2 = .737$, and an interaction of both factors, $F(2.1, 59.5) = 63.195$, $p < .001$, $\eta p^2 = .685$ (see Figure 5A). Responses were fastest for correct detections (2.2 sec), intermediate for misplaced highlights and misses (3.0 and 3.1 sec, respectively), and very slow when faults were absent for correct rejections and false alarms (3.6 and 3.6, respectively). All pairwise comparisons were significant, all $p$s < .001, except for the ones between misplaced highlights and misses, $p = .369$, and between correct rejections and false alarms, $p > .9$. Participants responded faster with XAI than BBAI (2.7 vs. 3.5 sec, respectively). However, just like in Experiment 1, XAI only improved performance for correct detections, correct rejections, and false alarms, all $p$s < .008. Also replicating Experiment 1, no significant XAI effects were found for misplaced highlights and misses, both $p$s > .07. All mean values are provided in Table 2.

*Error rates.* The ANOVA revealed a main effect of (X)AI accuracy, $F(2.0, 58.6) = 88.032$, $p < .001$, $\eta p^2 = .752$. The main effect of XAI availability was not significant, $F(1,29) = .041$, $p = .842$, $\eta p^2 = .001$, but XAI availability interacted with (X)AI accuracy, $F(2.9, 83.7) = 36.487$, $p < .001$, $\eta p^2 = .557$ (see Figure 5B). The main effect of (X)AI accuracy indicated that first, error rates were highest when a fault was present but not correctly indicated (41.7 and 47.0 % for misplaced highlights and misses, respectively), with the error rates in these conditions being about twice as high as in Experiment 1. Second, error rates were low when faults were absent (2.5 and 8.1 % for correct rejections and false alarms, respectively), while these numbers were rather similar to Experiment 1. Third, overall error rates for correct detections (17.3 %) were somewhere in between those of the other conditions. All pairwise comparisons between (X)AI accuracy conditions were significant, $p < .02$, except for the one between misplaced highlights and misses, $p = .753$. The absence of a main effect of XAI availability indicated that XAI did not improve performance in general when averaged across the five (X)AI accuracy conditions (23.5 vs. 23.2 %, respectively), but the presence and direction of XAI effects varied with (X)AI accuracy. For correct detections, XAI led to a major reduction of error rates (from 32.7 to 1.9 %), $p < .001$, whereas for misplaced highlights, it led to a large increase (from 29.3 to 54.1 %), $p < .001$. For the remaining (X)AI accuracy conditions, XAI effects were not significant, all $p$s > .09.

**Figure 5**

Performance results for Experiment 2, depending on XAI availability and (X)AI accuracy. (A) Mean reaction times and (B) error rates. Error bars represent standard errors of the mean.

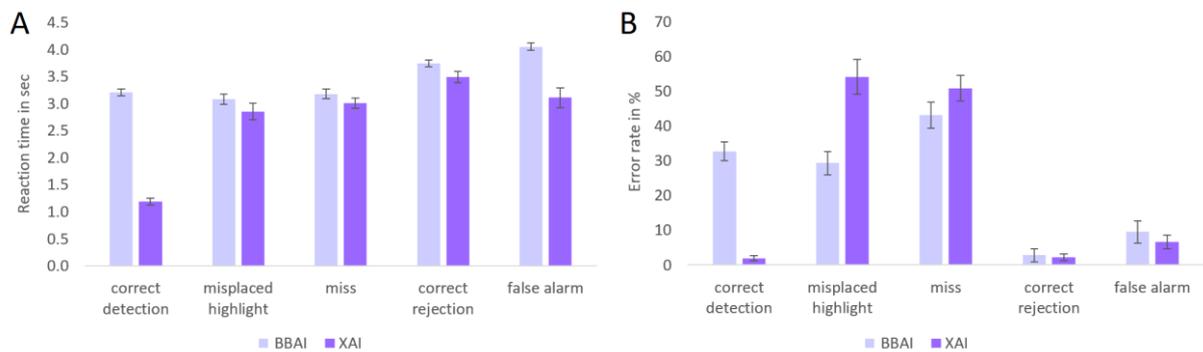

*3.2.2 Eye Movements*

*Dwell times within different AOIs*. The times participants spent fixating different AOIs are presented in Figure 6A. For misplaced highlights, they spent less time searching the rest of the mould with XAI than BBAI (1.6 vs. 2.6 sec), $t(29) = -6.510$, $p < .001$, $d = 1.189$. The same was found for false alarms, with participants spending only about half as much time inspecting the mould with XAI as compared to BBAI (2.1 vs. 3.9 sec), $t(29) = -10.684$, $p < .001$, $d = 1.951$.

*Fixations within the fault AOI*. We again analysed the percentage of trials in which participants fixated the fault AOI, using a 3 (*(X)AI accuracy: correct detection, misplaced highlight, miss*) x 2 (*XAI availability: XAI, BBAI*) repeated measures ANOVA. There was a main effect of (X)AI accuracy, $F(2,58) = 31.507$, $p < .001$, $\eta p^2 = .521$, and while there was no main effect of XAI availability, $F(1,29) = .049$, $p = .826$, $\eta p^2 = .002$, it interacted with (X)AI accuracy, $F(1.6,46.1) = 43.702$, $p < .001$, $\eta p^2 = .601$ (see Figure 6B). The fault AOI was fixated more often for correct detections than misplaced highlights and misses (83.7 vs. 64.2 and 68.7 %), both $p$s < .001, while misplaced highlights and misses did not differ, $p = .308$. For correct detections, participants more often fixated the fault with XAI than BBAI (93.9 vs. 73.5), $p < .001$, while the reverse was true for misplaced highlights (53.6 vs. 74.9 %), $p < .001$. For misses, fault fixations did not differ between XAI and BBAI (68.4 and 68.9), $p = .887$.

*Individual strategies*. Did the increased task difficulty of Experiment 2 affect participants' individual strategies of search in misplaced highlight trials? First, the combination of decision outcome and evidence revealed that most participants still made justified decisions in the majority of trials (see

Figure 6C). However, the number of non-justified decisions was higher than in Experiment 1, both for "okay" decisions without having fixated the rest of the mould, and "faulty" decisions without having fixated the fault. Second, the time participants spent searching the mould again varied considerably between participants, and only few participants refrained from searching almost completely (see Figure 6D). The variation between participants seemed even more gradual than in Experiment 1.

**Figure 6**

Eye movement results for Experiment 2. (A) Dwell times within different AOIs, (B) Trials with fixations within the fault AOI. The data for correct rejections and false alarms appear in grey as no faults were visible. Error bars represent standard errors of the mean. The last two parts of the figure depict individual participants' search behaviour for misplaced highlights: (C) Trialwise combination of decision and evidence, (D) Time spent searching the rest of the mould

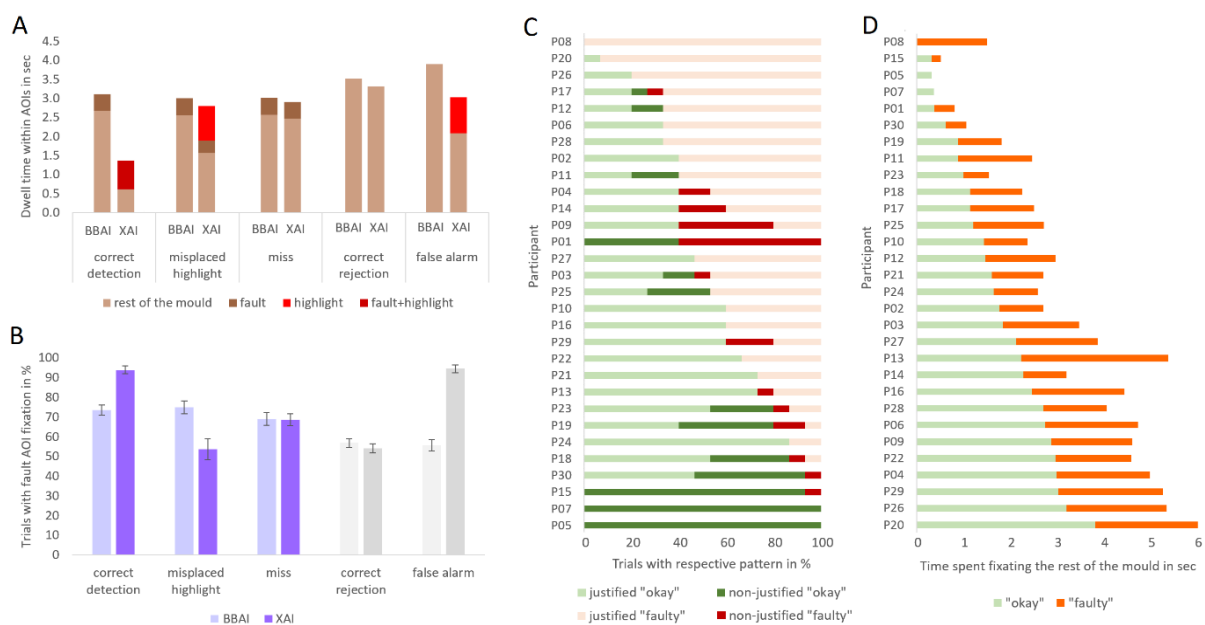

### 3.3 Discussion

In Experiment 2, we aimed to replicate the results of Experiment 1 while eliminating its methodological problems, increasing the number of relevant trials and using a uniformly high task difficulty. Indeed, the task turned out to be more difficult, as indicated by a 10 % increase in error rates compared to Experiment 1 for BBAI. In consequence, the effects of XAI (both benefits and costs) were much higher as well, in line with the finding that people more heavily rely on visual cues in difficult tasks (Maltz & Shinar, 2003). Still, we replicated all relevant results of Experiment 1. Not surprisingly, for correct detections XAI speeded up reactions and reduced errors again (in line with H1-a and H1-b). More interestingly, we also replicated the XAI reaction time benefit for correct rejections (supporting H2-a but not H2-ab0). Thus, the capability of an XAI system to explain its decisions can affect performance even when no highlights are present, which might be due to a halo or reassurance effect (Alberdi et al., 2004). It is worth noting that in this earlier study, the cueing benefit for correct rejections appeared in error rates, whereas in our two experiments it only affected reaction times, probably because our stimulus material makes commission errors unlikely.

It remains striking that again we did not find a comparable effect for misses. When the AI incorrectly decided that the mould was intact, its general capability to generate explanations did not significantly increase omission errors (contradicting H3-b). This was also evident in eye movements, where for misses the mould was not fixated less long and the fault was not fixated less often with XAI than BBAI. Thus, participants apparently did search for faults in the XAI condition, and no less thoroughly. The puzzle why XAI effects in highlight-absent trials existed for correct rejections but not for misses might be solved by looking at the absolute dwell time values on the rest of the mould (see Figure 6A). Even the dwell times for correct rejections, which were supposedly cut short by XAI, were more than 750 ms longer than the respective dwell times for misses. It seems like a potential "reassurance effect" was prevented because by the time when it could have kicked in, participants had already found the fault in half of the miss trials. In other words, if the XAI capabilities of our AI system indeed gave participants reassurance, this only made the difference between a long and a very long search time.

Most importantly, we replicated the effects of XAI that highlighted an intact area. For false alarms, such XAI speeded up performance (supporting H4-a), as participants seemed to quickly decide that no fault was present, which enabled responses to be similarly correct as they were without XAI (supporting H4-b0). For misplaced highlight, the same shortcutting of search substantially increased error rates (in line with H5-b). Although again XAI did not speed up responses for misplaced highlights (refuting H5-a), it led to a shorter search phase on the rest of the mould, resulting in fewer fixations of the fault compared to BBAI.

A look at the interindividual differences in eye movements for misplaced highlights suggests that the proportion of non-justified decisions increased compared to Experiment 1 (more dark areas in Figure 6C than 4C). This suggests that the higher task difficulty urged participants to make decisions based on less evidence. It also is in line with research suggesting that a higher task load, either due to verification complexity, time pressure, or multitasking requirements, reduces cross-checking behaviour (Lyell & Coiera, 2017; Manzey et al., 2014; Rieger & Manzey, 2022). However, again these non-justified decisions were a minority, and also the time spent searching varied gradually between participants. This is interesting, because one could have argued that a higher task difficulty would make the adoption of a distinct no-search strategy more attractive.

The gradual variation in search time was accompanied by a gradual variation in performance. Reaction times and error rates for misplaced highlights spanned almost the entire value range, and participants who often indicated that no fault was present were rather quick in doing so. Conversely, there are no indications that any participant blindly trusted the XAI, which ironically should have led to very fast and accurate performance in case of misplaced highlights. One might wonder why the same correlation did not appear in Experiment 1. We suspect this to be a consequence of variations in task difficulty due to the different fault types. Taken together, it seems worthwhile to consider the mechanisms behind observable XAI effects on performance, and we will take this up in the General Discussion.

## 4 General Discussion

Explainable artificial intelligence (XAI) has obvious potentials to support humans in visual inspection tasks. However, research on visual cueing has repeatedly shown that cueing benefits go hand in hand with cueing costs when cues are incorrect. Do these cueing costs generalise to the application of XAI in domains such as visual quality control, where observers have to ensure that products do not deviate from a clearly defined standard? And if so, under what conditions? The present study investigated how the impacts of XAI on visual inspection performance and processes depend on the accuracy of AI decisions and XAI highlights. Participants had to check chocolate moulds for faults and were supported by a simulated AI system that either acted like a black box or was able to explain its decisions by highlighting relevant image areas.

We found XAI costs only under one particular circumstance: when an image was correctly classified as faulty by the AI, but this decision was incorrectly explained by a misplaced XAI highlight. In this case, participants seemed to take the XAI highlight for granted, prematurely concluding from it that the AI decision was wrong. Accordingly, they tended to contradict this decision, instead of thoroughly checking whether it might be based on other image areas. We also found evidence that this XAI-induced performance cost for misplaced highlights was due to reduced visual search, as participants spent less time scanning the rest of the mould and thus fixated the fault less often. However, large interindividual differences suggested that overall XAI effects were based on gradual variations between participants, rather than discrete and clearly distinguishable strategies.

To our knowledge, the present study is the first one that relates behavioural cueing effects to the underlying search processes as revealed by eye movements. This allowed us to clarify how some of our behavioural effects came about (e.g., why XAI increased error rates but did not reduce reaction times for misplaced highlights, what participants actually did after seeing a highlight on an intact area, whether this was consistent across the participant sample). Understanding these search processes is of theoretical importance, because it can differentiate between different mechanisms that could bring about one and the same behavioural result. Throughout the following sections, we will integrate our discussion of XAI effects with explanations of the underlying search processes inferred from our eye movement data. We will first discuss the benefits and costs of XAI, critically examining whether participants overrelied on it. Specifically, we will focus on the selective cost for misplaced highlights and ask whether they really deteriorated performance more than any other type of AI error. Moreover, we will contrast our findings with previous research that identified false alarms as being most detrimental to human performance. We will conclude by making the limitations of the present study transparent and outlining directions for future research.

### 4.1 Benefits of XAI in Visual Quality Control

When pooled across all trials, XAI enabled faster and more accurate performance. In principle, it is not surprising to find XAI benefits, as the same has been shown for direct visual cues in various studies over the years (e.g., Chavaillaz et al., 2018; Maltz & Shinar, 2003; Warden et al., 2023). What is more surprising is that these performance benefits occurred despite our AI system's relatively low reliability of 70 %. According to a review of the automation literature, with an accuracy below 70 % automated aids tend to lose their benefits, and aided performance drops below the level of manual performance (Wickens & Dixon, 2007). In line with this general influence of system reliability, visual cueing benefits in detection performance often only emerged when the cues were highly reliable (e.g., Yeh & Wickens, 2001). The divergence of the present findings from this literature does not stem from the fact that we used black box AI (BBAI) and thus indirect cueing as a baseline, instead of manual performance. As Experiment 1 also included manual blocks, we were able to conduct a control analysis to compare error rates between all three levels of support: manual, BBAI, and XAI. This analysis revealed XAI benefits compared to both other conditions, which did not differ from each other.

This raises the question why our XAI was beneficial overall, despite its low reliability. Apparently, neither misses nor false alarms had serious negative impacts on performance, in contrast to their effects in previous studies. In the following sections, we will try to solve this puzzle step by step. But first, let us take a broader look at the cueing effects of our XAI highlights. A cueing or XAI system that has clear benefits will typically go along with a high risk of overreliance (Maltz & Shinar, 2003; Warden et al., 2023; Yeh & Wickens, 2001). Therefore, in the next section we will address the question whether the present study provided any evidence for overreliance on the (X)AI.

### 4.2 Were There Any Signs of Overreliance on (X)AI?

The performance benefits we found for correct AI were offset by performance costs or null effects for incorrect (X)AI. Does this mean that participants overrelied on the XAI? Overall, we did not find much evidence of overreliance. If anything, one observation could be interpreted as a sign of high trust, namely that participants were faster in agreeing with the AI's correct rejections when the system had XAI capabilities, although no highlights were shown (cf. Alberdi et al., 2004). However, there are two arguments against an explanation in terms of overreliance. First, the effect was small, and search times still were higher than in most other conditions. Second, if the system's XAI capabilities had led participants to blindly accept its decision when highlights are absent, this should also have manifested in higher omission error rates for misses. In principle, overtrust could also have kept participants from noticing it when the highlights pointed to an intact area. This should have manifested in fast responses, combined with commission error rates for false alarms and highly accurate performance for misplaced highlights. None of this was found. In sum, overreliance did not appear to be a problem in the present study. Again, we need to ask whether this might have been a consequence of our AI system's low reliability, which can not only reduce cueing benefits but also cueing costs (Maltz & Shinar, 2003; Yeh & Wickens, 2001). In the section on false alarms, we will explain why we do not think this is the case.

Conceptually, it is important to consider that it cannot be judged whether people overrely on an automated system without defining what amount of information sampling would be optimal (Moray, 2003; Moray & Inagaki, 2000). Extensive cross-checking is not always beneficial (Manzey et al., 2014). Across a broad range of psychology studies, participants tend to prioritise accuracy over speed and seem to have an almost moral obligation to avoid errors, even when this is refuted by objective payoffs (Fiedler & McCaughey, 2023). This tendency becomes problematic in the light of findings that people often deteriorate overall system performance by interfering with the automation in an attempt to improve it (Boskemper et al., 2022). Obviously, the right balance between relying on a system and critically cross-checking its outcomes depends on the respective work domain. It can be questioned whether it is a good idea to spend large amounts of time and effort on finding the one in several hundred chocolate bars that an AI system might have missed. Or as Dixon and colleagues have eloquently put it (Dixon et al., 2013, p. 459): "quarantining a portion of our attention for unexpected events is wasteful in most circumstances."

### 4.3 How Did Participants Deal with Misplaced Highlights?

The only XAI cost we found was that for misplaced highlights. But what does it mean, conceptually? It seems like participants' verification procedures were cut short by the XAI, leading them to only cross-check the highlight but not check the rest of the mould as thoroughly. This is in line with the finding that automated decision support diminishes people's search activities, although not leading to a complete lack of search (Rieger & Manzey, 2022). However, the rationality of this approach can be questioned. Taking an empty highlight as evidence that the AI decision is wrong and the opposite is true (i.e., the chocolate is okay) might not strike readers as overly irrational in the current scenario. However, the same conclusion would not have appeared all that rational in another, analogous classification task. Imagine participants did not have to distinguish faulty from intact chocolate but dogs from other animals. The mere fact that the XAI points to a non-informative area (e.g., grass in the background) would not have made participants conclude that the image is not a dog. Rather, they would probably have assumed that either the AI or the XAI does not work properly.

One explanation of our participants' behaviour and the resulting XAI costs for misplaced highlights might be that participants were not sufficiently aware of the functioning of XAI. Accordingly, they might not have considered that the link between AI decisions and XAI explanations is not deterministic, and that right decisions can be wrongly explained (Müller, Thoß, et al., 2023). A practical conclusion might be that XAI users should be made more aware of its limitations. However, a second potential explanation is that participants were not irrational but simply economic. A crucial difference between our chocolate scenario and the dog classification analogy is their verification complexity. When it is a major effort to verify the correctness of AI decisions plus XAI highlights, participants might simply have adopted a more heuristic approach. That is, they might have based their assessment of the AI's decisions merely on the evidence that it claims to have used, rather than on all other potential evidence that it might have used but that was hard to collect. In this case, a practical conclusion might be to critically examine the development, design, and application of XAI, instead of debiasing people.

A final explanation attempt starts from the critical question whether wrong explanations really were the culprit. Did they actually have more detrimental effects than all other ways in which an (X)AI system can err? At first glance, our selective XAI costs for misplaced highlights seem at odds with previous findings that cueing costs were similar for misplaced highlights and misses (Alberdi et al., 2004; Goh et al., 2005). But are our results really different from those findings? A closer look at our error rates for misplaced highlights and misses shows that they did not actually differ in the XAI condition. Thus, it did not matter whether the XAI provided a wrong highlight or no highlight at all. The marked difference in XAI costs between misplaced highlights and misses was in fact due to the BBAI condition in Experiment 2. With BBAI, misplaced highlights were perceptually identical to indirectly cued correct detections (and were only classified as "misplaced highlights" to allow for fair statistical comparisons between parallel trials). In other words, when participants used an AI system without XAI capabilities and saw a fault-present image, they were more inclined to falsely conclude that it was intact when the AI claimed this (i.e., misses, 43.1 % errors) than when the AI made them aware that the image was faulty (i.e., "misplaced highlights", 29.3 % errors). Accordingly, misplaced highlights were no more problematic than misses, their cueing costs simply resulted from being compared to a higher level of baseline performance. In practical terms, this tells us that false highlights resulting from bad XAI methods may sometimes have more adverse effects than showing no highlights at all, particularly when the AI decisions are likely correct. Explainability in general and visual cueing in particular may not be the best choice under all circumstances (cf. Boskemper et al., 2022; Maltz & Shinar, 2003; Yeh & Wickens, 2001).

### 4.4 Why Did We Not Find Cueing Costs for False Alarms?

There is a striking difference between the present study and many previous ones. We did not find XAI costs for false alarms, while previous studies often came to the opposite conclusion, namely that false alarms have particularly adverse effects, much more so than misses (Rice & McCarley, 2011; Wickens et al., 2005). Sometimes this was attributed to cues merely changing observers' criterion (i.e., making them more liberal) rather than improving their sensitivity, and thus promoting commission errors (Maltz & Shinar, 2003; Yeh & Wickens, 2001). In contrast, our commission error rates for false alarms were quite low, and did not increase with XAI. Presumably, this can be traced back to our stimulus material. In visual quality control, the aim is to ensure that products conform to a clearly defined standard and do not show any deviations. Thus, there were no non-problematic deviations acting as distractors, and no other obstacles to target identification (e.g., clutter, different perspectives, superimposition). Thus, the main challenge was to decide whether a deviation was present, not whether it was a target or not. This is not to say that our stimulus material was easy to work with, as evidenced by the high omission errors for misses in the BBAI condition (i.e., 43.1 % in Experiment 2). However, the challenging question was not *what* was cued (i.e., identification, including decision-making) but merely *whether* the cued location contained an object (i.e., localisation). In other words, verification effort was rather low once a fault was found. This is also supported by our participants' dwell times on the fault, which were highly consistent across conditions, suggesting that no prolonged scanning and decision-making was needed.

This also leads us back to the issues addressed in the previous sections, namely why in contrast to other studies we found XAI benefits despite the low AI reliability, but rarely any XAI costs. Due to the more complex stimuli used in previous studies, their cues presumably did not only affect participants' search processes but also their decision process. It has been argued that such effects may be inevitable, despite not being intended by system designers (Alberdi et al., 2008). For instance, the presence of a cue for a false alarm might convince observers that an ambiguous deviation is suspicious after all, whereas the absence of a cue for a miss might reassure them that it is not. None of this seemed to be an issue in our visual quality control task.

**4.5 Limitations and Future Research**

Several factors limit the conclusions to be drawn from the present study, and raise questions that might pave the way for future research. A first set of limitations concerns factors that created artificial conditions for the sake of convenience and experimental control. For instance, we relied on a student sample, whereas cueing effects are known to depend on domain expertise (Hättenschwiler et al., 2018). Our AI system was simulated and generated a fixed number of decisions for each (X)AI accuracy condition, whereas real AI system usually are designed to be more prone to false alarms than misses. Depending on the XAI method, misplaced highlights could also be much more or less prevalent.

A second threat to external validity is our high signal rate, or base rate of faults (i.e., 66 %), which presumably made participants expect moulds to be faulty. Thus, they may have adopted other strategies if signal rate had been as low as it is in most practical settings (although faults can in fact occur with a high frequency in the food industry, Müller & Oehm, 2019). However, the question whether high signal rates are realistic does not only depend on the domain, but also the task definition. If observers' job is to review exactly those pre-selected images for which the AI was uncertain, a high signal rate might even be the norm. Still, future studies should examine how our findings generalise to situations with lower signal rates. For instance, the gradual variations in search times for misplaced highlights might make room for more discrete strategies when cross-checking turns out to be futile again and again. On the other hand, it might not, as people are reluctant to refrain from cross-checking even when the signal rate is extremely low (Manzey et al., 2014).

A third limitation concerns our stimulus material. For one, the XAI highlights were visualised as orderly, uniform bounding boxes, perfectly centred around the fault area. This might be considered a surface feature, but cue visualisation can influence cueing effects in general and XAI effects in particular (Krupinski et al., 1993b; Sundararajan et al., 2019). Moreover, although our chocolate moulds were realistic, their characteristics limit the generalisation of our results. We already discussed the issue of stimulus complexity in the section on false alarms. Additionally, our faults per se were small defects restricted to one location. During real chocolate production, a wide range of qualitatively different faults is observed. First, sometimes several locally restricted faults are present in one mould. What if XAI occasionally highlighted some of them but missed others? This was investigated in another recent study (Müller, Mangold, et al., 2023). Resembling the phenomenon of satisfaction of search or subsequent search misses (Adamo et al., 2021), a small subset of participants strategically changed their inspection behaviour: they refrained from searching for additional faults when one fault was highlighted. Second, deviations in chocolate moulds can subtend large areas of several bars, for instance when the filling shines through with different degrees of severity. How might XAI change observers' criterion in this context, when the task is not to find an isolated fault but to decide whether the deviation is severe enough to warrant discarding the mould? There is preliminary evidence that visual cues can change the way people evaluate how critical an ambiguous deviation actually is (Alberdi et al., 2008). We are currently planning to systematically investigate this question.

Finally, an interesting challenge for future research will be to specify and differentiate the cognitive mechanisms that underlie cueing effects in different domains. Currently available studies on cueing and XAI differ immensely in the cognitive functions that their cues are supposed to support, which may not always go hand in hand with the functions they actually support (Alberdi et al., 2008). This calls for an explicit consideration of domain specificity in cueing studies, combined with systematic process tracing methodologies such as eye tracking to compare visual inspection processes across domains. As an implication for practice, it appears that we cannot easily generalise across domains, and should take domain characteristics into account when designing and using XAI.